\documentclass[aps,prd,onecolumn,floatfix,nofootinbib,superscriptaddress]{revtex4-1}
\usepackage[colorlinks,linkcolor=blue,anchorcolor=blue,citecolor=blue,urlcolor=blue,breaklinks=true]{hyperref}
\usepackage[libertine]{newtxmath}
\usepackage{graphicx}
\usepackage{slashed}
\usepackage{amsfonts}
\usepackage{amsmath}
\usepackage{blindtext}
\usepackage{microtype}
\usepackage{easyReview}
\usepackage{booktabs}
\usepackage{cleveref}
\usepackage{bm}
\usepackage{extarrows}
\usepackage[font=small]{caption}
\usepackage{subcaption}
\usepackage{float} 

\newcommand{\vect}[1]{\boldsymbol{#1}}

\def\ua{\uparrow}
\def\da{\downarrow}

\begin{document}

\title{Diffractive electroproduction of light vector particles: leading Fock-state contribution in the presence of significant higher Fock-state effects.}

\author{Chao Shi}
\email[]{cshi@nuaa.edu.cn}
\affiliation{Department of Nuclear Science and Technology, Nanjing University of Aeronautics and Astronautics, Nanjing 210016, China}

\author{Liming Lu}
\affiliation{Department of Nuclear Science and Technology, Nanjing University of Aeronautics and Astronautics, Nanjing 210016, China}

\author{Jian-feng Li}
\email[]{ljf169@ntu.edu.cn}
\affiliation{School of Physics and Technology , Nantong University, Nantong 226019, China}

\author{Wenbao Jia}
\affiliation{Department of Nuclear Science and Technology, Nanjing University of Aeronautics and Astronautics, Nanjing 210016, China}

\begin{abstract}

We study exclusive diffractive production of vector mesons and photon using the color dipole model with leading Fock-state light-front wave functions derived from Dyson–Schwinger and Bethe–Salpeter equations. New results for the $\phi$ meson and real photon are presented. Without data fitting, our calculation well matches HERA data in certain kinematical domains. The key finding of this paper is that in a color dipole model study for $\rho/\gamma$ and $\phi$, where light quarks are involved, the leading $q\bar{q}$ approximation is valid only when $Q^2$ exceeds $20$ and $10~\text{GeV}^2$ respectively, unlike $J/\psi$ which can be well described for $Q^2\approx 0$ GeV$^2$. This underscores the special role of $\phi$ electroproduction in color dipole picture: it strikes a balance between the large dipole size typical of light mesons and the smaller size associated with high-$Q^2$ photons, making it potentially well-suited for probing gluon saturation effects.
\end{abstract}
\maketitle

\section{INTRODUCTION\label{intro}}

Exclusive diffractive electroproduction of vector particles such as real photon and vector mesons constitutes an important probe of the transverse gluon density in hadrons and its saturation at small Bjorken‑$x$ \cite{Nemchik:1996pp,Nemchik:1996cw,Golec-Biernat:1999qor,Forshaw:2003ki,Fucilla:2024yfl}. In the color dipole approach description, the incoming virtual photon fluctuates into a quark–antiquark  dipole and then transforms back into an  outgoing vector particle, both described by their $q\bar{q}$ light‑front wave functions ($q\bar{q}$-LFWFs). Therefore the light vector particle production, such as $\rho$, $\phi(s\bar{s})$ and $\gamma$ has long been facing great challenge due to the large uncertainty in their $q\bar{q}$-LFWFs.

Although many existing studies achieve good agreement with data employing phenomenological light vector meson wave function models \cite{Nemchik:1996pp,Nemchik:1996cw,Forshaw:2003ki,Kowalski:2006hc,Forshaw:2012im,Rezaeian:2013tka,Roa:2023skv,Gurjar:2024wpq,Gurjar:2025kcp}, the connection between these models and realistic quantum chromodynamics (QCD) remains elusive. A primary problem, from our perspective, is that these wave functions assume the $|q\bar{q}\rangle$ state saturates the meson state. The normalization condition $\langle q\bar{q}|q\bar{q}\rangle = 1$ is usually imposed, so there is no (or little) higher Fock-state content such as $|q\bar{q}g\rangle$ in light mesons. This is a rough approximation, given the complex partonic structure of light hadrons. Meanwhile, recent next-to-leading-order calculations provide infrared‑safe analytic expressions for exclusive $\rho$ and $\phi$ production containing both $|q\bar{q}\rangle$  and $|q\bar{q}g\rangle$  contributions, yielding good agreement with HERA and LHC data \cite{Mantysaari:2022bsp,Fucilla:2023mkl}. Notably, therein the $|q\bar{q}g\rangle$ is obtained by perturbative emission of gluon from $|q\bar{q}\rangle$ at large $Q^2$, hence a suppression can be expected.

The light vector particle electroproductions receive particular interest as they were believed to be more sensitive to gluon saturation effects than heavy quarkonium such as $J/\psi$. This is due to the large dipole size of the light mesons that overlap with the flavor-independent $q\bar{q}$-nucleon scattering amplitude. Searching the saturation effect with light vector meson production in nuclei-nuclei and electron-nuclei collisions is thus appealing \cite{Goncalves:2017wgg,Armesto:2014sma,Accardi:2012qut}. On the other hand, HERA data on light vector particle electroproduction serves as a complement to the heavy meson case. It is important to utilize these data to further constrain or examine theoretical understanding of nonperturbative QCD quantities as $q\bar{q}$-LFWFs and/or color-dipole-nucleon scattering amplitudes.
 
In \cite{Shi:2021taf}, we introduced a light front projection method that extracts the $\rho$ and $J/\psi$ meson $q\bar{q}$-LFWFs from their covariant Bethe-Salpeter (BS) wave functions based on Dyson-Schwinger equations (DSEs) formalism. These $q\bar{q}$-LFWFs were then put into the color dipole study of the mesons' diffractive electroproduction. The key finding therein is that in $\rho$ meson the $|q\bar{q}\rangle$ contributes less than 50\%, i.e., $\langle q\bar{q}|q\bar{q}\rangle<0.5$, implying significant higher Fock-states contribution. Further more, the exclusive $\rho$ electroproduction can be well described with our $q\bar{q}$-LFWFs within color dipole approach, but only starting from $Q^2\approx 10$ GeV$^2$. This is reasonable in physics as twist suppression requires high $Q^2$ in exclusive processes. Nevertheless, we note this scenario is demonstrated for the first time within color dipole approach as other studies all describe data for $Q^2$ as low as 1 GeV$^2$. For $J/\psi$, our study suggested $\langle q\bar{q}|q\bar{q}\rangle \approx 0.9$, and the $q\bar{q}$-truncated color dipole model well describes data for $Q^2 \approx$ 0 GeV$^2$. In this work, we supplement \cite{Shi:2021taf} with cases of $\phi$ meson and real photon, which are all light vector particles. The novelty of this work is thus a first presentation of  $\phi$ meson $q\bar{q}$-LFWFs based on Dyson-Schiwnger equations formalism, as well as an exploration of the exclusive production of $\phi$ and $\gamma$ using DS-BSEs based $q\bar{q}$-LFWFs. We will also present a more detailed and more careful analysis on differential cross sections, with focus on light quark sector. We note that the nonperturbative $q\bar{q}$-LFWFs of real photon are adopted from \cite{Shi:2023jyk}, which is also based on DSEs formalism but with a simpler interaction model than the Maris-Tandy like model \cite{Shi:2021taf} for vector mesons.  

This paper is organized as follows. In section \ref{sec:LF-LFWF}, we recapitulate $q\bar{q}$-LFWFs of vector mesons and photon, and give the result of $\phi$. In section \ref{sec:DVMP}, these $q\bar{q}$-LFWFs were utilized in a color dipole model study of  vector mesons and photon electroproduction. We finally conclude in section \ref{sec:con}.  

\section{$q\bar{q}$-LFWFs of vector mesons and photon \label{sec:LF-LFWF}}

\subsection{Formalism}
A particle state takes a Fock-state expansion on the light front. For vector meson $V_M$ composed of valence quark and antiquark of flavor $f$, the decomposition reads
\begin{align}
|V_M\rangle &=|q_f\bar{q}_f\rangle_{(V_M)}+|q_f\bar{q}_f g\rangle_{(V_M)}+...\label{eq:FockExpand1}
\end{align}
On the other hand, in the context of QCD plus quantum electrodynamics (QED), the photon Fock-state expansion reads  
\begin{align}
|\gamma^*_{\rm phys} \rangle &=|\gamma^*_{\rm bare}\rangle+|e^+e^-\rangle_{(\gamma^*)}+\sum_{f=u,d,s...}|q_f\bar{q}_f\rangle_{(\gamma^*)}+\sum_{f=u,d,s...}|q_f\bar{q}_fg\rangle_{(\gamma^*)}+...\label{eq:FockExpand2}
\end{align}
Every term on the right hand side of Eqs.~(\ref{eq:FockExpand1}, \ref{eq:FockExpand2}) shares exactly same quantum number as its parent particle state, and is mutually orthogonal to the each other. The photon is an elementary particle and thus its Fock-state expansion contains an additional bare term.  Meanwhile, its $q\bar{q}-$LFWFs run through all quark flavors. At a first glance, the idea that the elementary particle photon has quark content may seem confusing, but will soon be clarified.

Denoting $V=\gamma^*$ and $V_M$, a general decomposition of the leading Fock-state $|q_f\bar{q}_f\rangle_{(V)}$ reads
\begin{align}
	|q_f\bar{q}_f\rangle^\Lambda_{(V)} &= \sum_{\lambda,\lambda';i,j}\int \frac{d^2 \vect{k}_T}{(2\pi)^3}\,\frac{dx}{2\sqrt{x\bar{x}}}\, \frac{\delta_{ij}}{\sqrt{3}} \Phi^{\Lambda,(f)}_{\lambda,\lambda',(V)}(x,\vect{k}_T)\, b^\dagger_{f,\lambda,i}(x,\vect{k}_T)\, d_{f,\lambda',j}^\dagger(\bar{x},\bar{\vect{k}}_T)|0\rangle. \label{eq:LFWF1}
\end{align}
The $\Phi^{\Lambda,(f)}_{\lambda,\lambda',(V)}$ is the $q\bar{q}$-LFWF of particle $V$ with helicity $\Lambda$ and quark (antiquark) of flavor $f (\bar{f})$ and spin $\lambda$ ($ \lambda'$). The $\Lambda=0, \pm 1$ and $\lambda=\ua$, $\da$, denoted as $\ua=+$ and $\da=-$ for abbreviation in following. The $b^+$ and $d^+$ are creation operators of quark and antiquark, respectively. The $i$ and $j$  are the color indices. The $\vect{k}_T=(k^x,k^y)$ is the transverse momentum of the quark, and $\bar{\vect{k}}_T=-\vect{k}_T$ for antiquark. The longitudinal momentum fraction carried by quark is $x=k^+/P^+$, with $\bar{x}=1-x$ for antiquark. We use the light-cone four vector convention $A^{\pm} = \tfrac{1}{\sqrt{2}}(A^0 \pm A^3)$ and $\vect{A}_T=(A^1, A^2)$ throughout this paper. 

By Eqs.~(\ref{eq:FockExpand1}-\ref{eq:LFWF1}), one can observe that the $q\bar{q}$-LFWFs are essentially transition amplitudes of parent particle $V$ into the quark-anti-quark state $b^\dagger_{f,\lambda,i}(x,\vect{k}_T)\, d_{f,\lambda',j}^\dagger(\bar{x},\bar{\vect{k}}_T)|0\rangle$.  So $q\bar{q}$-LFWFs of photon should not be viewed as photon's bound state wave function, but rather the transition amplitude of photon into a virtual $q\bar{q}$ pair by quantum fluctuation. Naturally, this interpretation also applies to vector mesons. It is also based on this idea that a connection can be built between the BS wave function which is the transition amplitude of $V\rightarrow q\bar{q}$ in ordinary space-time coordinate, and $q\bar{q}$-LFWFs based on light-front coordinate \cite{tHooft:1974pnl,Liu:1992dg,Carbonell:1998rj}. In \cite{Shi:2021taf}, we introduced a light front projection equation to obtain $q\bar{q}$-LFWFs of vector mesons from their BS wave functions, i.e., 
\begin{align}
	\Phi^{\Lambda,(f)}_{\lambda,\lambda',(V)}(x,\vect{k}_T)&=-\frac{1}{2\sqrt{3}}\int \frac{dk^- dk^+}{2 \pi} \delta(x Q^+-k_\eta^+)\textrm{Tr}\left \{\Gamma_{\lambda,\lambda'}\gamma^+ S_f(k_\eta)\left[\Gamma^{(f)}_{(V)}(k;Q)\cdot \epsilon_\Lambda(Q)\right]S_f(k_{\bar{\eta}})  \right\}.\label{eq:chi2phi}
\end{align}
The $S_f(k)$ is the fully dressed quark propagator of flavor $f$, and $\Gamma_{(V),\mu}^{(f)}(k;Q)$ is the amputated $V\rightarrow q_f\bar{q}_f$ vertex, i.e., the BS amplitude, in the momentum space. The BS wave function is defined as $\chi_{(V),\mu}^{(f)}(k;Q)\equiv S_f(k_\eta)\Gamma_{(V),\mu}^{(f)}(k;Q)S_f(k_{\bar{\eta}})$. The $Q$ is the four momentum of vector meson. The $k_\eta\equiv k+\eta Q$ is the momentum carried by outgoing quark leg (corresponding to quark content) and $k_{\bar{\eta}}\equiv k-(1-\eta)Q$ is that carried by antiquark. The $\epsilon^\mu_\Lambda(Q)$ is the polarization vector for vector particle. Choosing $\Gamma_{\pm,\mp}=I\pm \gamma_5$ or $\Gamma_{\pm,\pm}=\mp(\gamma^1\mp i\gamma^2)$ \footnote{The convention is that the $+$ and/or $-$ signs in the same row should be simultaneously taken in one equation. This convention is adopted throughout this paper.} can project out a LFWF with specific quark-antiquark helicity configuration. In color space there is a unit matrix on the right hand side of Eq.~(\ref{eq:chi2phi}). The trace is taken in Dirac and color spaces. 

Due to various symmetry constraints, the $\Phi^{\Lambda }_{\lambda,\lambda'}(x,\vect{k}_T)$'s can further be expressed with five independent scalar amplitudes $\psi(x,\vect{k}_T^2)$'s \cite{Carbonell:1998rj,Ji:2003fw,Shi:2021taf}, i.e.,
\begin{align}
	\hspace{00mm}\Phi_{\pm,\mp}^{0}&=\psi^{0}_{(1)},  \ \ \ \ \ 
	&\Phi_{\pm,\pm}^{0}&=\pm k_T^{(\mp)} \psi^{0}_{(2)}, \label{eq:phi1}\\
	\Phi_{\pm,\pm}^{\pm 1}&=\psi^{1}_{(1)},
	&\Phi_{\pm,\mp}^{\pm 1}&=\pm  k_T^{(\pm)}\psi^{1}_{(2)}, \notag \\
	\Phi_{\mp,\pm}^{\pm 1}&=\pm k_T^{(\pm)}\psi^{1}_{(3)},
	&\Phi_{\mp,\mp}^{\pm 1}&=(k_T^{(\pm)})^2\psi^{1}_{(4)}. \label{eq:phi2}
\end{align}
with $k_T^{(\pm)}=k^x \pm i k^y$ and $\psi^{1}_{(2)}(x,\vect{k}_T^2)=-\psi^{1}_{(3)}(1-x,\vect{k}_T^2)$. They form a complete set of possible $q\bar{q}$-LFWFs of unflavored vector meson and virtual photon. For real photon, i.e., the limit $Q^2\rightarrow 0$ of virtual photon, only transverse components exists hence $\psi^0_{(i)}$'s are all zero. 

It is also customary to classify the $q\bar{q}$-LFWFs by their quark-anti-quark orbital angular momentum (OAM) projected along the $z$-axis, denoted by $l_z$. The $l_z$ can be determined by $\Lambda=\lambda+\lambda'+l_z$ due to angular momentum conservation. Given all possible spin configurations in $\Phi^{\Lambda,(f)}_{\lambda,\lambda'}$, the $l_z$ can be $0$, $\pm 1$ and $\pm 2$, and are referred to as s-, p- and d-wave components in the literature. The $l_z$ can also be read directly from the power of $k_T^{(\pm)}$ in Eqs.~(\ref{eq:phi1},\ref{eq:phi2}) \cite{Ji:2003fw}. For example, from the last equation of Eqs.~(\ref{eq:phi2}) we read $\Phi_{+,+}^{- 1}=\left(k_T^{(-)}\right)^2\psi^{1}_{(4)}$, hence the $l_z$ is negative due to minus sign in $k_T^{(-)}$, and there are two units of OAM due to the power of $2$, which preserves angular momentum conservation as both quark and antiquark are spin-up.

 \subsection{Vector meson $q\bar{q}$-LFWFs} 
\begin{figure}[htbp]    \includegraphics[width=0.4\linewidth]{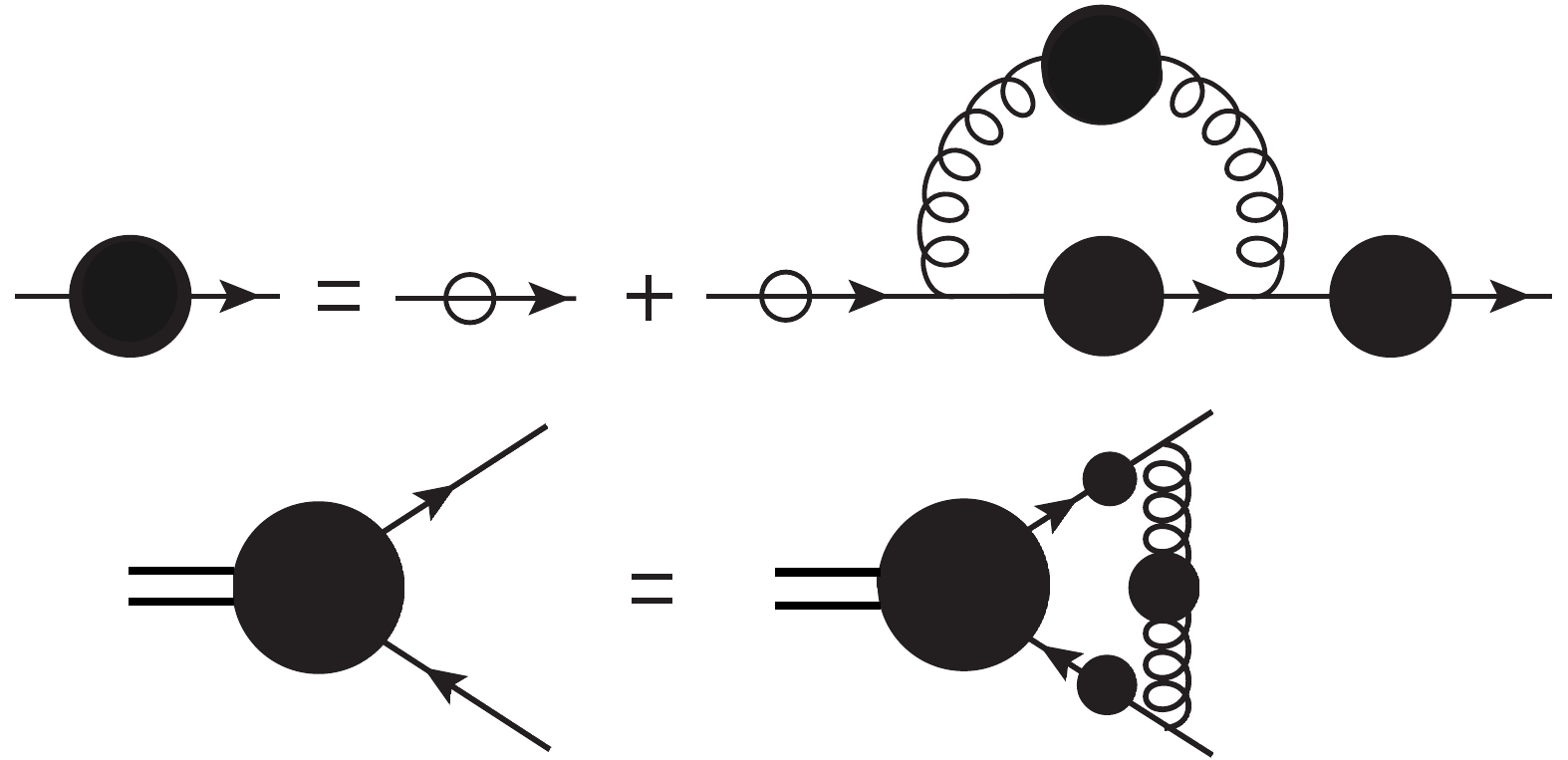}
    \caption{The Dyson-Schwinger equation of quark's dressed propagator (upper plot), and vector meson's Bethe-Salpeter equation of BS amplitude (lower plot) under rainbow-ladder truncation. The open circle is bare (perturbative) quark propagator and black blobs are dressed quark and gluon propagators.}\label{fig:VBSE}
\end{figure}

In \cite{Shi:2021taf}, we obtained the $q\bar{q}$-LFWFs of $\rho$ and $J/\psi$ with Eq.~(\ref{eq:chi2phi}). Therein the dressed quark propagator and meson BS amplitude are obtained by simultaneously solving the quark's Dyson-Schwinger and meson's BS equations, see Fig.~\ref{fig:VBSE} for a diagrammatic representation under the rainbow-ladder truncation. We employed a Maris-Tandy-like gluon propagator model \cite{Maris:1999nt,Qin:2011xq} and physical current quark mass for quarks. In this paper, we employ exactly same model and truncation setup for the DS and BSEs. For the $\phi$ meson to be considered in this paper,  the only one change made is to change current quark mass from $m_{u/d}=5$ MeV for $\rho$ meson to $m_{s}=95$ MeV for $\phi$  meson. After solving the DS and BSEs, the $\phi$ meson mass is solved to be 1.01 GeV and the leptonic decay constant is 0.176 GeV, close to PDG data 1.02 GeV and 0.170 GeV respectively \cite{ParticleDataGroup:2024cfk}. The $\phi$'s $q\bar{q}-$LFWFs are then extracted based on Eq.~(\ref{eq:chi2phi}), using technique that had been developed and explained with detail for $\rho$ in \cite{Shi:2021taf}.

In Fig.~\ref{fig:psis}, we show the s-wave component $q\bar{q}-$LFWFs of $\phi$, as compared to $\rho$ meson's. There are significant differences between the two. The $\phi$ meson $q\bar{q}-$LFWFs are generally narrower than $\rho$ in longitudinal momentum fraction $x$. This is in line with the finding that $q\bar{q}-$LFWFs with heavier quarks are generally more centered around $x=1/2$, i.e., in heavier mesons the quark and antiquark tend to share the longitudinal momentum equally. In the transverse momentum $\vect{k}_T$, the $\phi$ meson $q\bar{q}-$LFWFs are more broadly distributed than $\rho$ meson, but not by much. This indicates in the coordinate space, the transverse size of $q\bar{q}$ dipole of $\phi$ meson is smaller than $\rho$ meson, also not by much. This suggests $\phi$ meson can also be a sensitive probe to saturation effects as $\rho$. 
 
 Generally speaking, the meson state is normalized, i.e., $\langle V_M|V_M\rangle=1.0$. In all existing color dipole model studies, this was used as a normalization condition for $q\bar{q}-$LFWFs, i.e., enforcing $\langle q\bar{q}|q\bar{q}\rangle=1.0$. In \cite{Shi:2021taf}, we demonstrated with a DS-BSE calculation that in $\rho$ the $\langle q\bar{q}|q\bar{q}\rangle\lesssim 0.5$. Therein we calculated 
\begin{align}
N_{\lambda,\lambda',(V_M)}^{\Lambda,(f)}&\equiv ^{\Lambda,\lambda,\lambda'}_{(V_M)}\!\!\langle q_f\bar{q}_f|q_f\bar{q}_f\rangle_{(V_M)}^{\Lambda,\lambda,\lambda'} \nonumber \\
&=\int_0^1 dx \int \frac{d \vect{k}_T^2}{2(2 \pi)^3}  |\Phi^{\Lambda,(f)}_{\lambda,\lambda',(V)}(x,\vect{k_T})|^2. \label{eq:N2}
\end{align}
The results for $\phi$ meson, along with $\rho$ and $J/\psi$ mesons, are summarized in  Table.~\ref{tab:N}. The $N_{HF}\equiv 1-\sum_{\lambda,\lambda'}N^\Lambda_{\lambda,\lambda'}$ indicates contribution from higher Fock-states. We notice in $\phi$ meson, the $N_{HF}$ is around 50\% and hence still significant.  Meanwhile, by comparing $\rho$, $\phi$ and $J/\psi$, one can notice the s-wave $q\bar{q}$-LFWFs' contribution increases with current quark mass, and the p- and d-wave components show opposite. The total effect is a reduction in the $N_{\textrm{HF}}$. This is consistent with the finding  that as the meson gets heavier, the higher Fock-states and high orbital angular momentum states all get  suppressed.

\begin{figure}[htbp]\label{fig:psis} 
  \centering
  \begin{subfigure}{0.45\textwidth}
    \includegraphics[width=\linewidth]{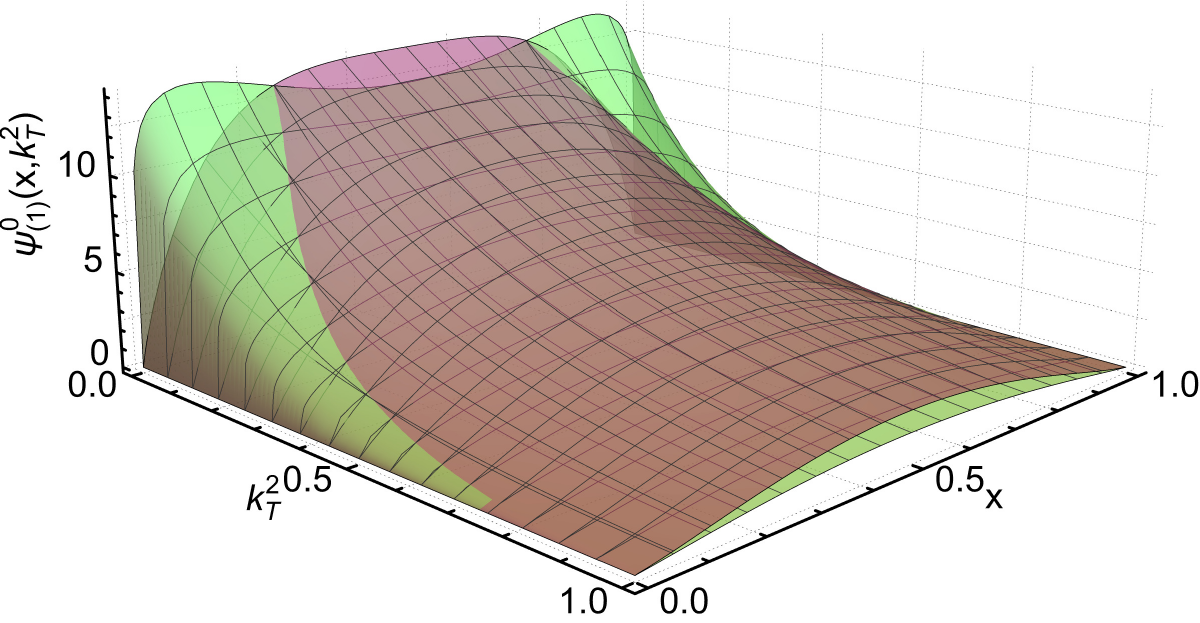}
  \end{subfigure}
  \hspace{0.5em}
  \begin{subfigure}{0.45\textwidth}
    \includegraphics[width=\linewidth]{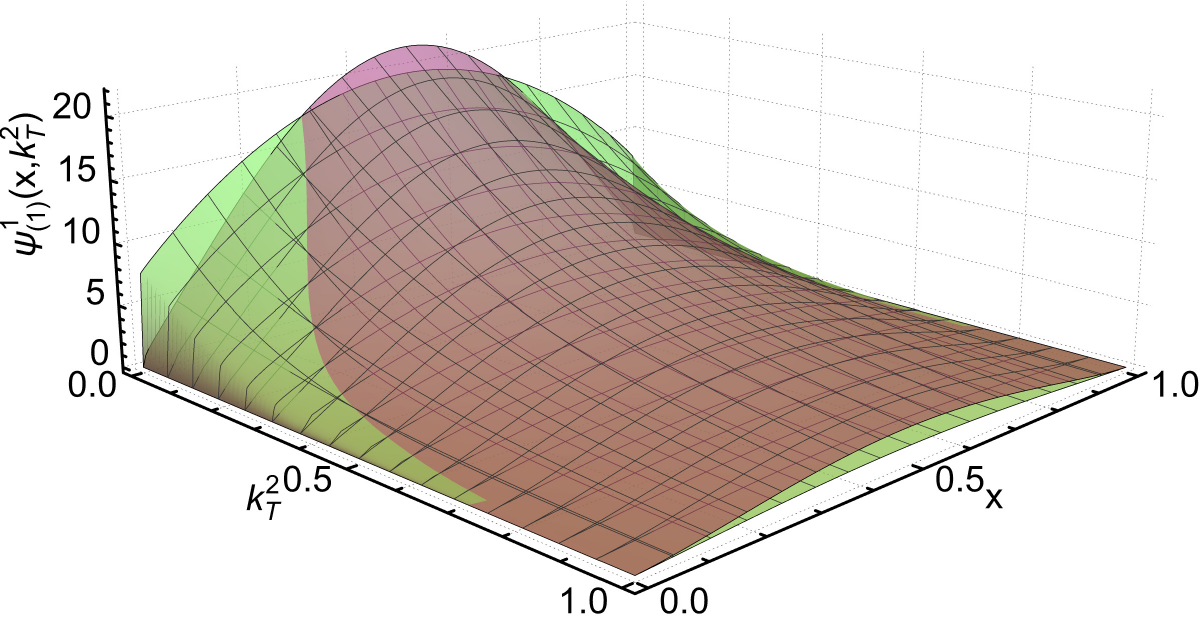}
  \end{subfigure}

  \caption{The dominant s-wave component $q\bar{q}$-LFWFs of $\phi$ (purple) and $\rho$ (green). See Eqs.~(\ref{eq:LFWF1},\ref{eq:phi1},\ref{eq:phi2}) for definition of the $\psi$'s.}
  \label{fig:psis}
\end{figure}

To conclude, we find that the $\phi$ meson $q\bar{q}$-LFWFs are different from $\rho$ in profile, but their $N_{\lambda,\lambda'}$'s are close. This reveals that $\phi$ meson is essentially a light meson, e.g., it is closer to $\rho$ instead of $J/\psi$. This is due to the fact that that the dynamical chiral symmetry breaking (DCSB) of QCD reduces the difference between u/d and s quarks. For instance, although the the current quark mass ratio is $m_s/m_u \approx 20$, after solving the quark's gap equation in Fig.~\ref{fig:VBSE}, the ratio of their quark mass functions is approximately $1.2$ by DCSB. The $\phi$ and $\rho$ mesons therefore host comparable internal soft dynamics, reflected by overall quantities such as $N_{\lambda,\lambda'}^{\Lambda}$'s. On the other hand, the $q\bar{q}$-LFWFs characterize detailed structural information of hadrons, and hence are sensitive to the current quark mass. Analysis on leading twist parton distribution amplitudes \footnote{The leading twist parton distribution amplitude is obtained by integrating over $\vect{k}_T$ in s-wave $q\bar{q}$-LFWFs.} of physical pion $\pi(u\bar{d})$ (130 MeV) and fictitious pion $\pi(s\bar{s})$ (690 MeV) had already shown significant discrepancy in their profile \cite{Shi:2021nvg}.

\begin{table}[h!]

\begin{center}
\begin{tabular*}
{\hsize}
{
l@{\extracolsep{0ptplus1fil}}
c@{\extracolsep{0ptplus1fil}}
c@{\extracolsep{0ptplus1fil}}
c@{\extracolsep{6ptplus1fil}}
c@{\extracolsep{0ptplus1fil}}
c@{\extracolsep{0ptplus1fil}}
c@{\extracolsep{0ptplus1fil}}
c@{\extracolsep{0ptplus1fil}}
c@{\extracolsep{0ptplus1fil}}
c@{\extracolsep{0ptplus1fil}}
c@{\extracolsep{0ptplus1fil}}}\hline
  & & $N_{\uparrow,\downarrow}$ & $N_{\downarrow,\uparrow}$ & $N_{\uparrow,\uparrow}$& $N_{\downarrow,\downarrow}$ & $N_{HF}$   \\\hline
$\Lambda=0$ & $\phi$ & 0.24 & 0.24 & 0.02 & 0.02  & 0.48  \\
& $\rho$ & 0.19 & 0.19 & 0.04 & 0.04  & 0.54  \\
&$J/\psi$ & 0.44 & 0.44 & 0.01 & 0.01  & 0.10  \\
$\Lambda=1$ & $\phi$ & 0.04 & 0.04 & 0.35 & 0.01  & 0.56 \\
&$\rho$ & 0.04 & 0.04 & 0.24 & 0.02  & 0.66  \\
&$J/\psi$ & 0.03 & 0.03 & 0.78 & $\approx 0.0$  & 0.16 \\\hline
\end{tabular*}
\end{center}
\vspace*{-4ex}
\caption{The $q\bar{q}$-LFWFs' contributions to Fock-states normalization. See Eq.~(\ref{eq:N2}) for definition of $N$. 
\label{tab:N}
}
\end{table}

\subsection{Photon $q\bar{q}$-LFWFs}
\begin{figure}[htbp] 
    \includegraphics[width=0.4\linewidth]{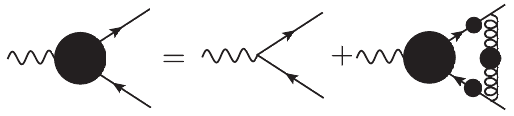}
    \caption{The photon's inhomogeneous Bethe-Salpeter equation of photon's BS amplitude $\gamma^*\rightarrow q\bar{q}$ under ladder truncation. }\label{fig:photonBSE}
\end{figure}

In \cite{Shi:2023jyk}, we utilized Eq.~(\ref{eq:chi2phi}) to obtain the photon's nonperturbative $q\bar{q}$-LFWFs.  The photon's BS wave function for $\gamma^*\rightarrow q\bar{q}$ is obtained by solving the contact interaction model within DSEs formalism. A diagrammatic representation for photon's BSE under ladder truncation is shown in Fig.~\ref{fig:photonBSE}. As compared to Fig.~\ref{fig:VBSE}, there is an additional bare term arising from leading order QED. The gluon exchange ladder is important for low virtuality $Q^2$, and gets suppressed at large $Q^2$, as a consequence of QCD's asymptotic freedom. It is therefore necessary to consider  nonpertubative QCD effect in photon with low virtuality, including real photon $Q^2=0$ GeV$^2$.

The extracted photon nonperturbative $q\bar{q}-$LFWFs are summarized as follows
\begin{align}
\psi_{(1)}^{0,(f)}(x,\vect{k}_T^2)&=e_f e P^{(f)}_T(Q^2)\frac{\sqrt{N_c}}{Q}\left(1-\frac{2x(1-x)Q^2}{k_\perp^2+\varepsilon_x^2}\right), \label{eq:psi01} \\
\psi^{0,(f)}_{(2)}(x,\vect{k}_T^2)&=0,\\
\psi^{1,(f)}_{(1)}(x,\vect{k}_T^2)&=e_f e P^{(f)}_T(Q^2)\sqrt{2N_c}\frac{M_f}{k_\perp^2+\varepsilon_x^2},\\
\psi^{1,(f)}_{(2)}(x,\vect{k}_T^2)&=e_f e P^{(f)}_T(Q^2)\sqrt{2N_c}\frac{x}{k_\perp^2+\varepsilon_x^2},\\
&=-\psi^{1,(f)}_{(3)}(1-x,\vect{k}_T^2),\\
\psi^{1,(f)}_{(4)}(x,\vect{k}_T^2)&=0. \label{eq:psi14}
\end{align}
with $\varepsilon_x\equiv \sqrt{Q^2 x(1-x)+M_f^2}$. The $P_T^{(f)}(Q^2)$ is a dressing function of quark-antiquark-photon vertex, i.e.,
\begin{align}
P^{(f)}_T(Q^2)&= \frac{1}{1+K^{(f)}_\gamma(Q^2)}, \\
K^{(f)}_\gamma(Q^2) &=  \frac{4\alpha_{\rm IR}M_f}{3\pi m_G^2}\int_0^1d\alpha\, \alpha(1-\alpha) Q^2\,  \overline{\cal C}^{iu}_1(\varepsilon_\alpha)\,
\end{align}
where ${\cal C}^{\rm iu}(M^2)/M^2 = \Gamma(-1,M^2 \tau_{\rm uv}^2) - \Gamma(-1,M^2 \tau_{\rm ir}^2)$, and $\Gamma(\alpha,y)$ is the incomplete gamma-function. Notations $\overline{\cal C}^{iu}_1(z)={\cal C}^{iu}_1(z)/z$ and ${\cal C}^{\rm iu}_1(z) = - z (d/dz){\cal C}^{iu}(z)$ are used. Model parameters involved are gluon mass $m_G = 0.5$ GeV, interaction strength $\alpha_{\rm IR}/\pi=0.36$, regulators $\Lambda_{\rm ir} = 0.24$ GeV and $\Lambda_{\rm uv} = 0.91$ GeV. We take current quark masses $m_{u/d}=0.007$ GeV and $m_s=0.095$ GeV, which produce $M_{u/d}=0.37$ GeV and $M_{s}=0.53$ GeV by  solving  quark's gap equation in Fig.~\ref{fig:VBSE}. 

In the color dipole model, the scattering amplitude is formulated in coordinate space, hence the photon LFWFs are Fourier transformed to coordinate space, i.e.,
\begin{align}
\tilde{\Phi}^{\Lambda,(f)}_{\lambda,\lambda',(V)}(x,\vect{r})=\int \frac{d^2\vect{k}}{(2\pi)^2}\textrm{e}^{i\vect{k}\cdot\vect{r}} \Phi^{\Lambda,(f)}_{\lambda,\lambda',(V)}(x,\vect{k}).\label{eq:Fourier}
\end{align}
Let $\boldsymbol r = \bigl(r\cos\theta_r,\,r\sin\theta_r\bigr)$,  Eqs.~(\ref{eq:psi01}-\ref{eq:psi14}, \ref{eq:Fourier}) yield
\begin{align}
\tilde{\Phi}^{0,(f)}_{\lambda,\lambda',(\gamma^*)}(\boldsymbol r,x;Q)
&=
\,e\,e_f\, P_T^{(f)}(Q^2) \frac{\sqrt{N_c}}{Q} \left( \delta^2(\vect{r})-2Q^2
\,x(1-x)\,\frac{K_0(\varepsilon_x\,r)}{2 \pi}
\;\right)\delta_{\lambda,-\lambda'} \label{eq:psir1}\\
\tilde{\Phi}^{\pm 1,(f)}_{\lambda,\lambda',(\gamma^*)}(\boldsymbol r,x;Q)&=
\,e\,e_f\, P_T^{(f)}(Q^2) \sqrt{2N_c}
\Biggl\{
\mp i\,e^{\pm i\theta_r}[x\,\delta_{\lambda,\pm}\,\delta_{\lambda',\mp}
      - (1-x)\,\delta_{\lambda,\mp}\,\delta_{\lambda',\pm}]\partial_{r}+ M_f\;\delta_{\lambda,\pm}\,\delta_{\lambda',\pm}\;\Biggr\}\frac{K_0(\varepsilon_x \,r)}{2\pi}.\label{eq:psir2}
\end{align}
The $\delta$-function term is always omitted in color dipole model study due to $\mathcal{N}(x,r=0,b)=0$, see Eq.~(\ref{eq:NbBCG}) below. Eqs.~(\ref{eq:psir1},\ref{eq:psir2}) only work at low $Q^2$, as the input photon BS wave function is solved using a low energy effective interaction model. For high $Q^2$, QCD effect gets suppressed and perturbative QED result can be used. The perturbative photon $q\bar{q}$-LFWFs can be obtained by making the replacements $P_T^{(f)}\rightarrow 1$ and $M_{f}\rightarrow m_f$ in Eqs.~(\ref{eq:psir1},\ref{eq:psir2}) \cite{Dosch:1996ss}. In \cite{Shi:2023jyk} we introduced a transition function to interpolate the results at low and high Q region,  
\begin{align}
\left|\tilde{\Phi}_{\textrm{Full}}\right|^2=F_{\rm part}(Q^2)\left|\tilde{\Phi}_{\textrm{NP}}\right|^2  +[1-F_{\rm part}(Q^2)]\left|\tilde{\Phi}_{\textrm{P}}\right|^2\label{eq:phiFull1}
\end{align}
The $\tilde{\Phi}$ is the abbreviation for $\tilde{\Phi}^{\Lambda,(f)}_{\lambda,\lambda',(\gamma^*)}(r,x;Q)$, with subscript NP for nonperturbative and P for perturbative. 
The transition function takes the parameterization
\begin{align}
F_{\rm part}(Q^2)=\frac{Q_0^{2n}}{(Q^2+Q_0^2)^n}.\label{eq:Fpart}
\end{align}
The determination of parameters $Q_0$ and $n$ will be explained below Eq.~(\ref{eq:B}).

For the purpose of exclusive vector particle production study in this paper, we further rewrite Eq.~(\ref{eq:phiFull1}) to get a direct expression for $\tilde{\Phi}_{\textrm{Full}}$. This is achievable, e.g., with 
\begin{align}
\tilde{\Phi}_{\textrm{Full}}=  \tilde{\Phi}_{\textrm{NP}}/\left|\tilde{\Phi}_{\textrm{NP}}\right|\sqrt{F_{\rm part}(Q^2)\left|\tilde{\Phi}_{\textrm{NP}}\right|^2  +[1-F_{\rm part}(Q^2)]\left|\tilde{\Phi}_{\textrm{P}}\right|^2}\label{eq:phiFull2}.
\end{align}
We employ Eq.~(\ref{eq:phiFull2}) to calculate every spin component of the interpolated full $q\bar{q}$-LFWFs.

\section{Exclusive vector meson and photon electroproduction within color dipole approach under $q\bar{q}$ truncation}\label{sec:DVMP}
\subsection{Formalism}
Diffractive vector-meson production is a class of exclusive processes in high-energy (virtual) photon–proton  collisions $\gamma^*p\rightarrow V_Mp$. Such event exhibits a large rapidity gap where no other particles are produced between vector meson $V_M$ and the intact target $p$. The vector meson can also be a real photon, namely the deeply virtual Compton scattering (DVCS). In the color dipole approach, the scattering amplitude for $\gamma^*p\rightarrow pV$ is written as an overlap of $\gamma^*$ and vector meson (or photon) $V$'s $q\bar{q}$-LFWFs convolved with the color-dipole–proton scattering amplitude $\mathcal{N}(x,r,b)$ \cite{Kowalski:2003hm,Hatta:2017cte}
\begin{align}
  \mathcal{A}_{T,L}^{\gamma^*p\rightarrow Vp}(x,Q,\Delta)
  \;=\;
  2i
  \int_{0}^{1} \frac{dz}{4\pi} 
  \int d^{2}\vect{r}\,
  \int d^{2}\vect{b}\;
  \bigl[\Psi_{(V)}^*\,\Psi_{(\gamma^*)}\bigr]_{T,L}
  \; e^{-\,i\,\bigl[\vect{b} - \left(\frac{1}{2}-z\right)\,\vect{r}\bigr]\cdot \vect{\Delta} }
  \;\mathcal{N}(x,r,b)\,.
\end{align}
Here $Q=\sqrt{-q^2}$ with virtual photon four momentum $q$. The $r = |\vect{r}|$ is the color dipole size, $b = |\vect{b}|$ is the impact parameter,  $\Delta =|\vect{\Delta}|$ is the momentum transfer between protons before and after scattering. The $x \;=\; \frac{Q^2 + M_V^2}{Q^2+W^2}$ is the longitudinal momentum fraction of proton carried by scattered gluon. The subscripts T and L represent transversely and longitudinally polarized vector particle, i.e., helicity $|\Lambda_V|=1$ or $\Lambda_V=0$ respectively. Note $\mathcal{A}_{L}=0$ for DVCS since the $q\bar{q}$-LFWFs vanish for longitudinal real photon. 

The overlap of $q\bar{q}$-LFWFs takes the form
\begin{align}
  \bigl[\Psi_{(V)}^*\,\Psi_{(\gamma^*)}\bigr]_{T}&=\frac{1}{2}\sum_{f=u,d,s,c}\ \sum_{\Lambda=\pm 1}\ \sum_{\lambda=\pm} \ \sum_{\lambda'=\pm}\left(\tilde{\Phi}_{\lambda,\lambda',(V)}^{\Lambda,(f)}(x,\vect{r})\right)^* \tilde{\Phi}_{\lambda,\lambda', (\gamma^*)}^{\Lambda,(f)}(x,\vect{r},Q)\label{eq:psioverlap1}\\
  \bigl[\Psi_{(V)}^*\,\Psi_{(\gamma^*)}\bigr]_{L}&=\sum_{f=u,d,s,c}\ \sum_{\Lambda=0}\ \sum_{\lambda=\pm} \ \sum_{\lambda'=\pm}\left(\tilde{\Phi}_{\lambda,\lambda',(V)}^{\Lambda,(f)}(x,\vect{r})\right)^* \tilde{\Phi}_{\lambda,\lambda', (\gamma^*)}^{\Lambda,(f)}(x,\vect{r},Q)\label{eq:psioverlap2}
\end{align}
Note when $V$ refers to a vector meson, the summation of flavor only covers the valence quarks. While when $V$ refers to real photon, the summation runs through all flavors. The real photon's $q\bar{q}$-LFWFs can be obtained by setting $Q=0$ in Eq.~(\ref{eq:phiFull2}). 

For the color-dipole-proton scattering amplitude ${\cal N}(x,r,b)$, we employ the impact parameter dependent color glass condensate (bCGC) model \cite{Watt:2007nr,Rezaeian:2013tka}, which reads 
\begin{eqnarray}
 {\cal N}(x,r,b)=\left\{
 \begin{aligned}
& N_0\left(\frac{r Q_s}{2}\right)^{2\gamma_{\rm eff}} \hspace{20 mm} rQ_s\le 2, \\ 
& 1-{\rm exp}[-{\cal A} {\rm ln}^2({\cal B}rQ_s)]  \hspace{5.2 mm} rQ_s >2,\label{eq:NbBCG}
 \end{aligned}
 \right.
\end{eqnarray}
with 
\begin{align}
Q_s(x,b)&=\left(\frac{x_0}{x}\right)^{\frac{\lambda}{2}} {\rm exp}\left[-\frac{b^2}{4\gamma_s B_{\rm CGC}}\right], \\
\gamma_{\rm eff}&=\gamma_s+\frac{1}{\kappa \lambda Y}{\rm ln}\left(\frac{2}{r Q_s}\right),  \\
Y&={\rm ln}(1/x),
\end{align}
and 
\begin{align}
{\cal A}&=-\frac{N_0^2 \gamma_s^2}{(1-N_0)^2{\rm ln}(1-N_0)},\\
{\cal B}&=\frac{1}{2}(1-N_0)^{-\frac{1-N_0}{N_0 \gamma_s}}.\label{eq:B}
\end{align}
The model parameters of bCGC model are directly employed from \cite{Shi:2023jyk}. The $\kappa=9.9$ and $B_{\rm CGC}=5.5$ GeV$^{-2}$ were chosen following \cite{Rezaeian:2013tka}, and the rest model parameters $N_0$, $\gamma_s$, $x_0$ and $\lambda$ are combined with $Q_0$ and $n$ in Eq.~(\ref{eq:Fpart}) to render a global fit to inclusive DIS reduced cross section data for $Q^2\in [0.25,50]$ GeV$^2$ \cite{Shi:2023jyk}. The best fit yields $N_0=0.4596$, $\gamma_s=0.6177$, $x_0=0.0001326$ and $\lambda=0.1875$, along with $Q_0^2=1.052$ GeV$^2$ and $n=3.97$. Note that the physical current quark masses $m_{u/d}\approx 0.005$ GeV, $m_s=0.095$ GeV and $m_c=1.27$ GeV were used throughout the calculation, without introducing phenomenological values such as $m_{u/d/s}=0.14$ GeV that had been popular in color dipole model studies \cite{Iancu:2003ge,Kowalski:2006hc,Albacete:2014fwa,Mantysaari:2018nng,Badelek:2022cgr,Dumitru:2023sjd}.

The differential cross section of diffractive vector particle electroproduction is
\begin{align}\label{eq:dsigmadt}
  \frac{d\sigma^{\gamma^*p\rightarrow Vp}_{L,T}}{dt}
  \;=\;
  \frac{1}{16\pi} 
  \bigl|\mathcal{A}_{L,T}^{\gamma^*p\rightarrow Vp}(x,Q^2,\Delta)\bigr|^2 (1+\beta^2) R_g^2,
\end{align}
The modification factor $1+\beta^2$ accounts for correction from imaginary part of $\mathcal{N}$, with $ \beta = \tan\!\Bigl(\tfrac{\pi}{2}\,\lambda\Bigr)$ and $\lambda= \frac{\partial  \ln(\mathcal{A}_{T,L}^{\gamma^*p\rightarrow Vp})}{\partial \ln(1/x)}$ \cite{Frankfurt:2000ez}. The $R_g(\lambda)=\frac{2^{\,2\lambda + 3}}{\sqrt{\pi}}\,\frac{\Gamma\bigl(\lambda + \tfrac{5}{2}\bigr)}{\Gamma\bigl(\lambda + 4\bigr)}$  is skewness factor which accounts for skewness effect when  gluons that interact with color dipole carries different momentum fractions \cite{Shuvaev:1999ce}.

\subsection{Results}

\begin{figure}[H] 
  \centering

  \begin{subfigure}{0.45\textwidth}
    \includegraphics[width=\linewidth]{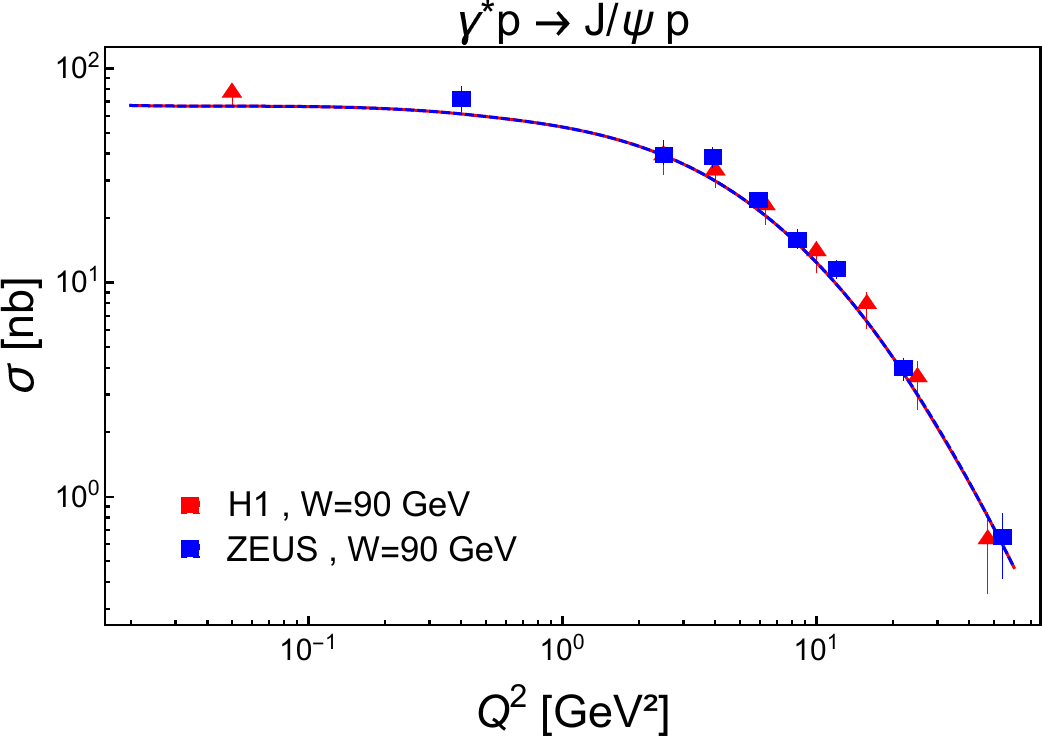}
  \end{subfigure}
  \hspace{0.5em}
  \begin{subfigure}{0.45\textwidth}
    \includegraphics[width=\linewidth]{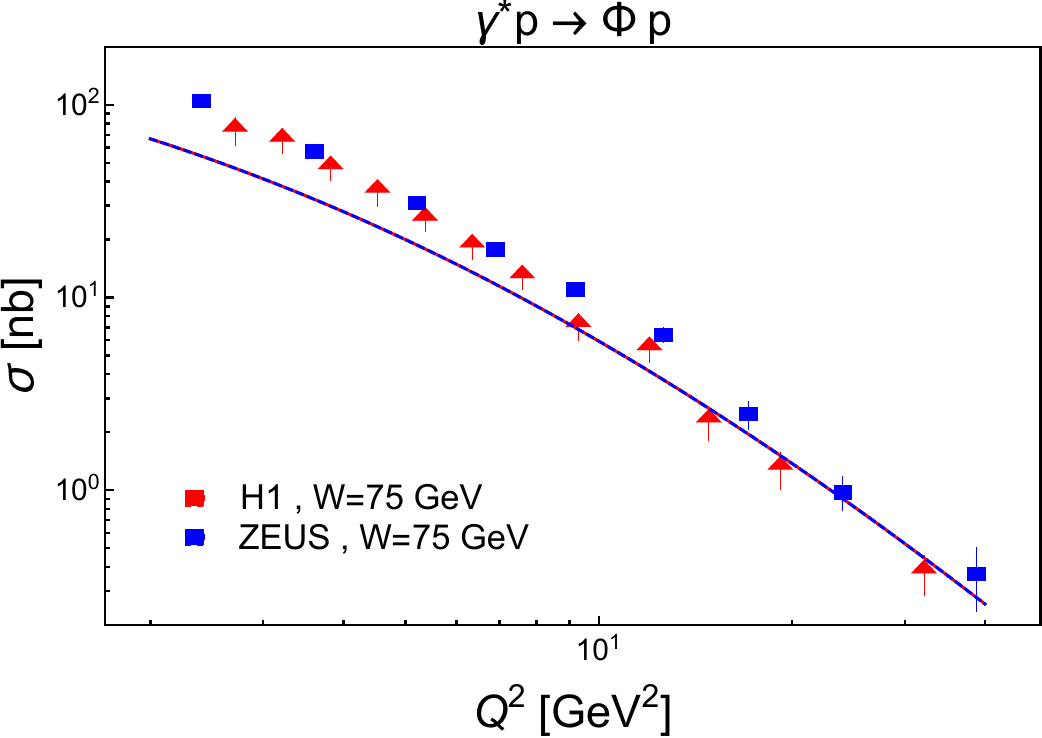}
  \end{subfigure}

  \vspace{1em}

  \begin{subfigure}{0.45\textwidth}
    \includegraphics[width=\linewidth]{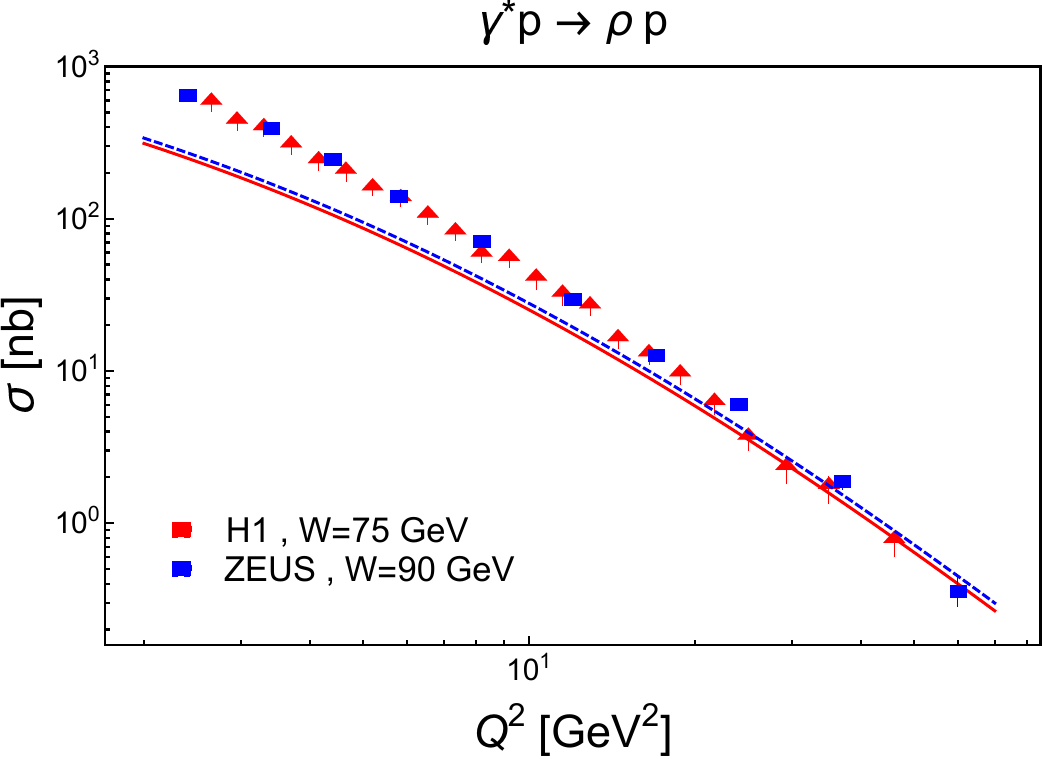}
  \end{subfigure}
  \hspace{0.5em}
  \begin{subfigure}{0.45\textwidth}
    \includegraphics[width=\linewidth]{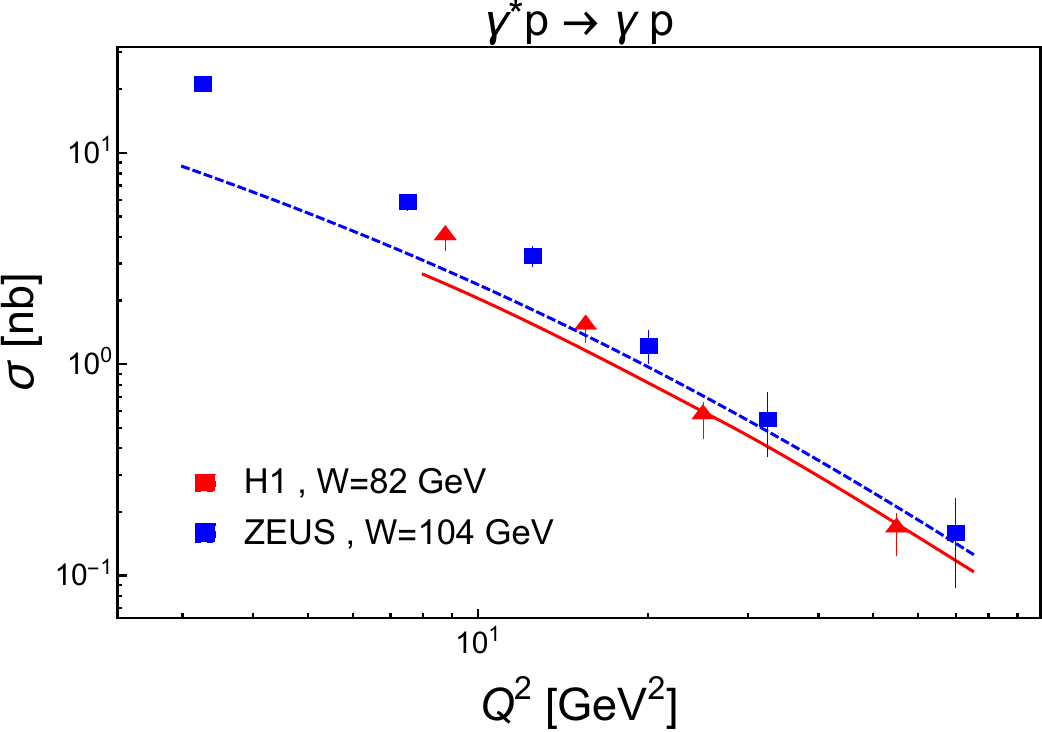}
  \end{subfigure}

  \caption{Total cross sections for $J/\psi$, $\phi$, $\rho$ and $\gamma$, as a function of $Q^2$. The red solid  and blue dashed curves are calculations to be compared with H1 and ZEUS data \cite{H1:2005dtp,ZEUS:2004yeh,H1:2009cml,ZEUS:2005bhf,ZEUS:2007iet,ZEUS:2008hcd,H1:2009wnw}.}
  \label{fig:sigQ2}
\end{figure}

In Fig.~\ref{fig:sigQ2} we show the cross section of exclusive electroproduction of light and heavy vector mesons and photon as a function of $Q^2$. For $J/\psi$, the result agrees well with data in almost the entire $Q^2$ range. As the meson gets lighter, the agreement shifts toward larger $Q^2$ region. Denoting $Q_c$ as the point where agreement starts, we find $Q_c^2\approx 10$, 20 and 20 GeV$^2$ for $\phi$ \footnote{For $\phi$ production, the ZEUS data is visibly larger than H1 data. Since our calculation aligns more closely with the H1 data, the conclusions are drawn based on comparisons with those results.}, $\rho$ and $\gamma$ respectively. This is reasonable, as in exclusive processes higher Fock-state contribution can only be suppressed by sufficiently high $Q^2$. Based on Table.~\ref{tab:N}, the leading Fock-state approximation works fine for $J/\psi$ since the $J/\psi$ is almost dominated by $q\bar{q}$ component, but for light mesons higher Fock-states can not be ignored unless the high $Q^2$  suppresses their contribution. Meanwhile, from Fig.~\ref{fig:sigQ2} we notice for $\phi$ meson the agreement starts from $Q_c^2 \approx 10$ GeV$^2$, which is lower than $\rho$ meson. Hence the $\phi$ meson is in a special position: On one hand, it can be more sensitive to saturation effects than $J/\psi$ for its large dipole size that is comparable with $\rho$. On the other hand, the leading Fock-state truncation works better for $\phi$ than $\rho$ and yields a $Q_c$ that is not too high. We note that an incoming photon with a higher $Q$ has a smaller dipole size, limiting  the size of overlapping color dipole in Eqs.~(\ref{eq:psioverlap1},\ref{eq:psioverlap2}), making the result less sensitive to saturation effects. In this sense, the $\phi$ meson exhibits a compromise between light $\rho$ meson and heavy $J/\psi$ meson, regarding the $Q_c$ and color dipole size. Finally, we notice for exclusive photon production, the $Q_c$ is close to $\rho$ meson's. This can be explained by noticing that the summation of photon $q\bar{q}$-LFWFs runs through all four flavors in Eqs.~(\ref{eq:psioverlap1},\ref{eq:psioverlap2}). This includes contribution from $u/d$ quarks, which is potentially populated with higher Fock-states at low scale.

\begin{figure}[H] 
  \centering

  \begin{subfigure}{0.45\textwidth}
    \includegraphics[width=\linewidth]{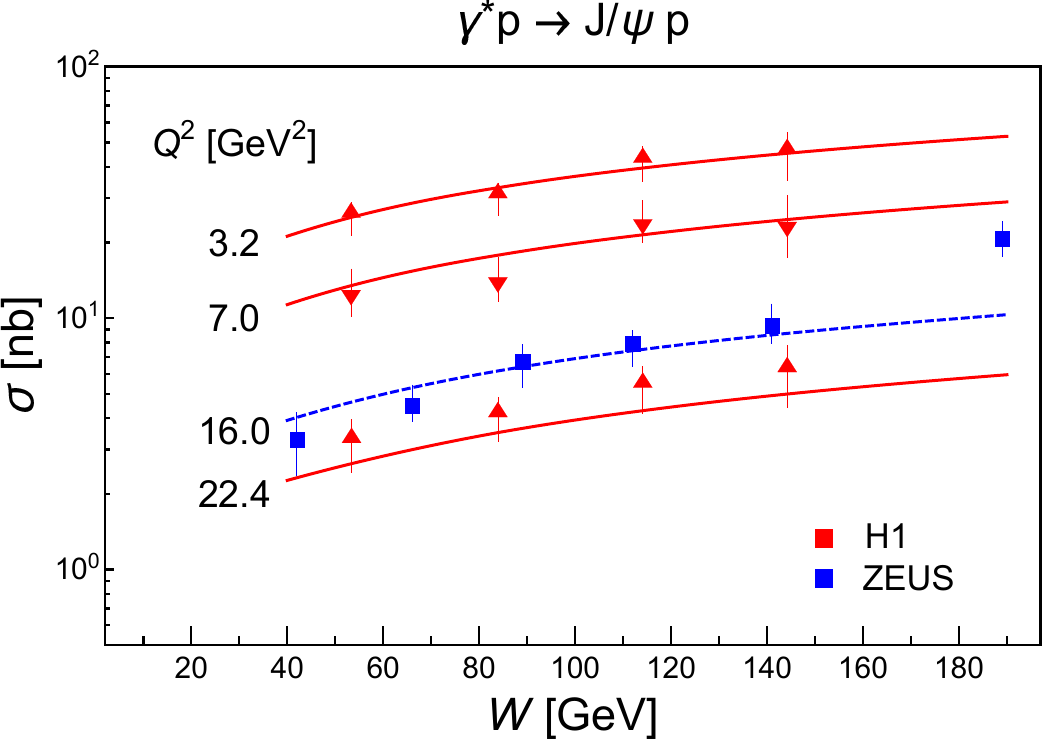}
  \end{subfigure}
  \hspace{0.5em}
  \begin{subfigure}{0.45\textwidth}
    \includegraphics[width=\linewidth]{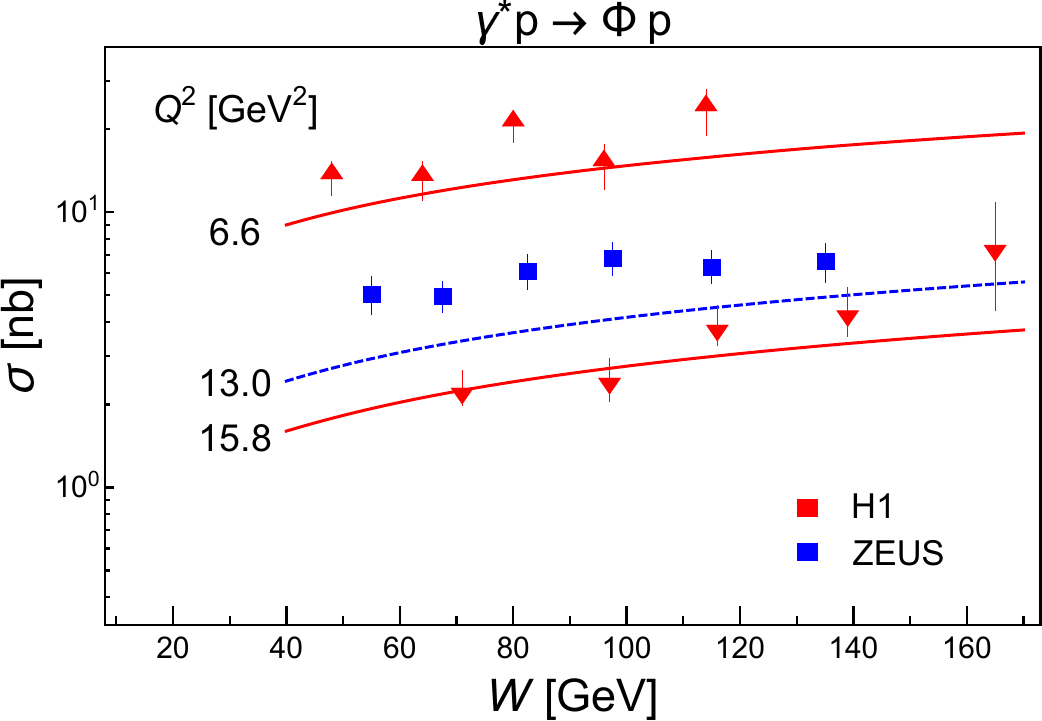}
  \end{subfigure}

  \vspace{1em}

  \begin{subfigure}{0.45\textwidth}
    \includegraphics[width=\linewidth]{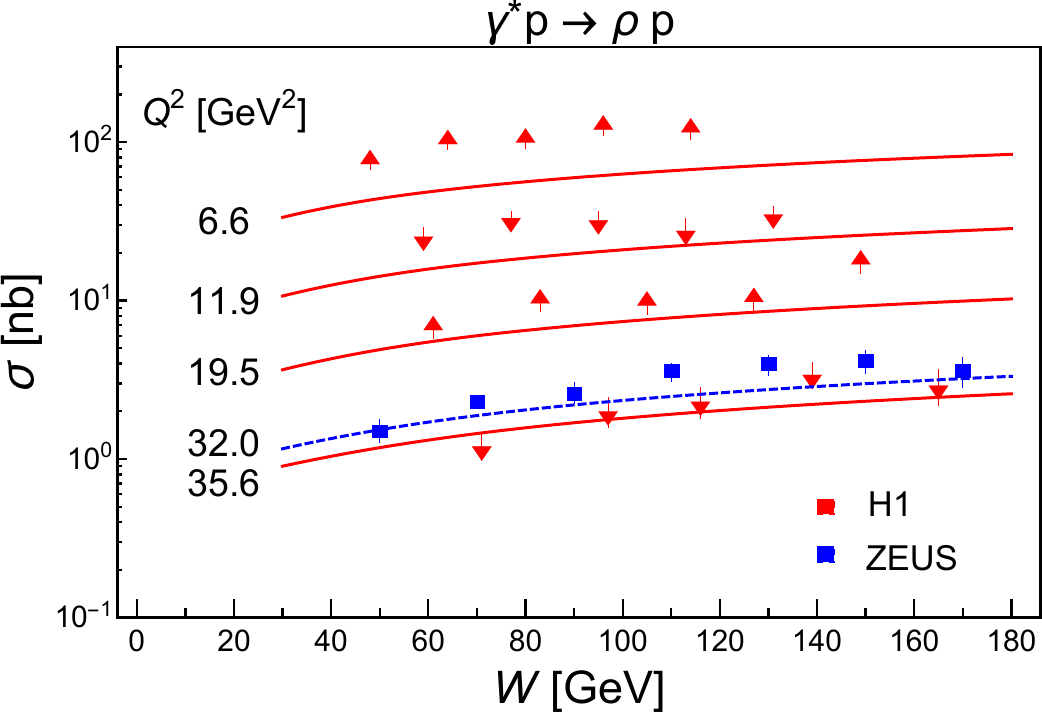}
  \end{subfigure}
  \hspace{0.5em}
  \begin{subfigure}{0.45\textwidth}
    \includegraphics[width=\linewidth]{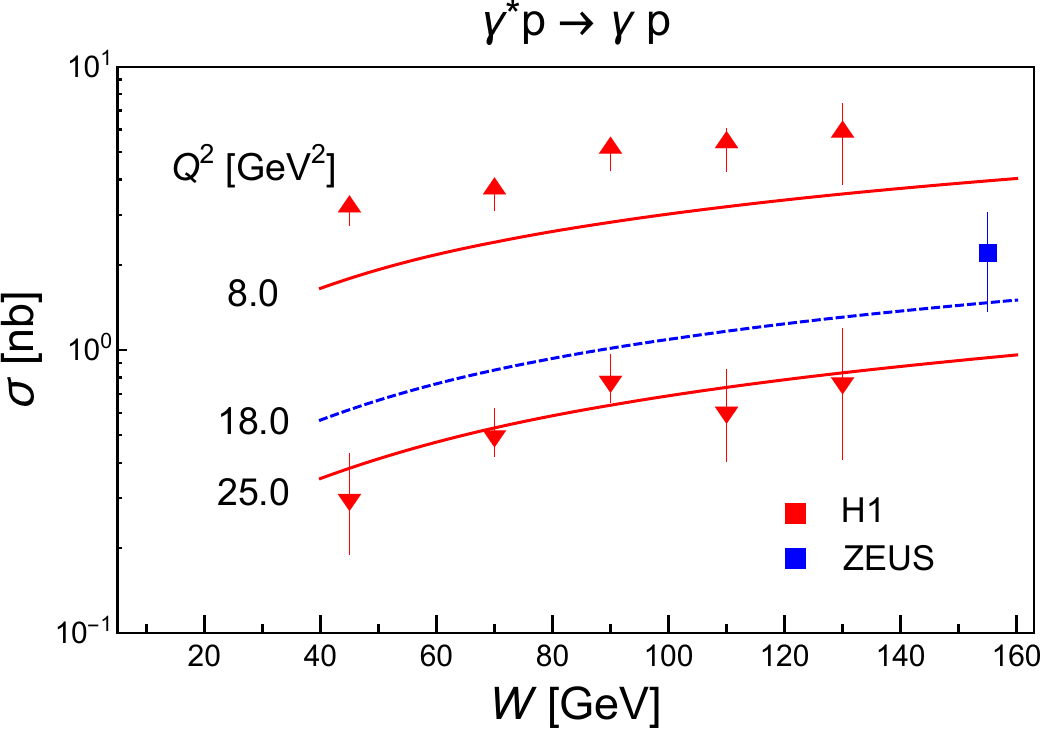}
  \end{subfigure}

  \caption{Differential cross sections for $J/\psi$, $\phi$, $\rho$ and $\gamma$, as a function of $W$. The red solid  and blue dashed curves are calculations to be compared with H1 and ZEUS data \cite{H1:2005dtp,ZEUS:2004yeh,H1:2009cml,ZEUS:2005bhf,ZEUS:2007iet,ZEUS:2008hcd,H1:2009wnw}.}
  \label{fig:sigW}
\end{figure}

 In Fig.~\ref{fig:sigW} the cross section as a function of $W$ is shown. The calculated $J/\psi$ cross sections agree well with data from low to high $Q^2$. For $\phi$ meson, our calculation shows a preference of H1 data for sufficiently large $Q^2=15.8$ GeV$^2$. The $Q^2=13.0$ GeV$^2$ data is larger than our calculation but we remind the ZEUS data is generally larger than H1 in this case. For $\rho$ meson, agreement shows up for $Q^2=19.5$ GeV. For DVCS, good agreement is found for $Q^2=18.0$ and $25.0$ GeV$^2$, but deviation is prominant for $Q^2=8.0$ GeV. All these results are in line with conclusion drawn from Fig.~\ref{fig:sigQ2}. 
 
\begin{figure}[H] 
  \centering

  \begin{subfigure}{0.45\textwidth}
    \includegraphics[width=\linewidth]{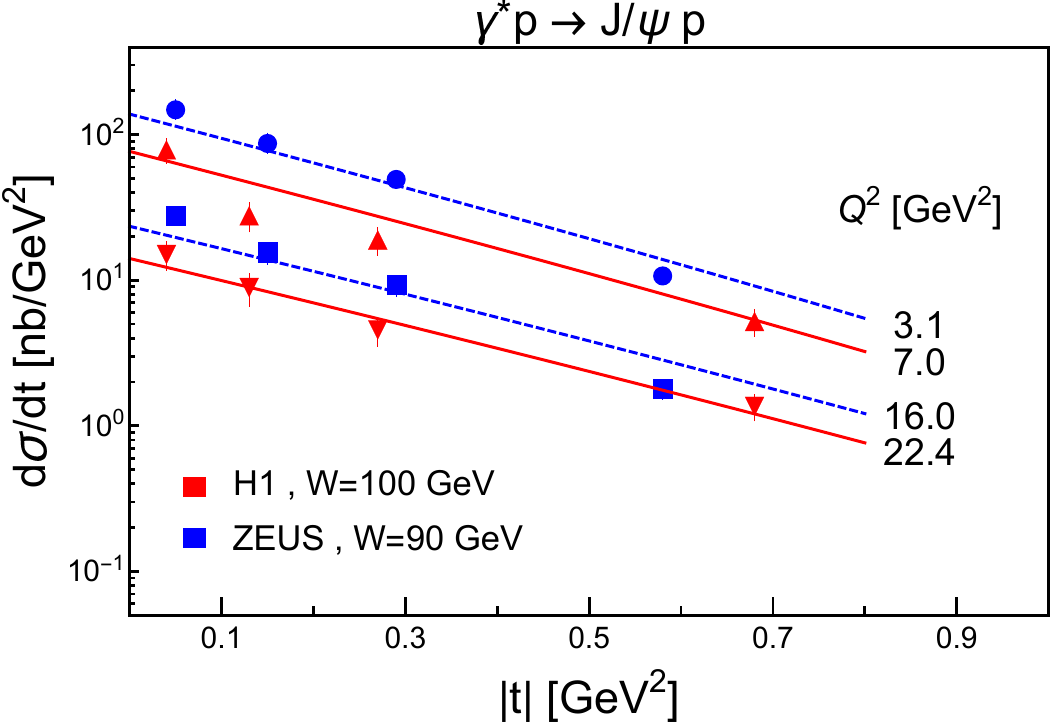}
  \end{subfigure}
  \hspace{0.5em}
  \begin{subfigure}{0.45\textwidth}
    \includegraphics[width=\linewidth]{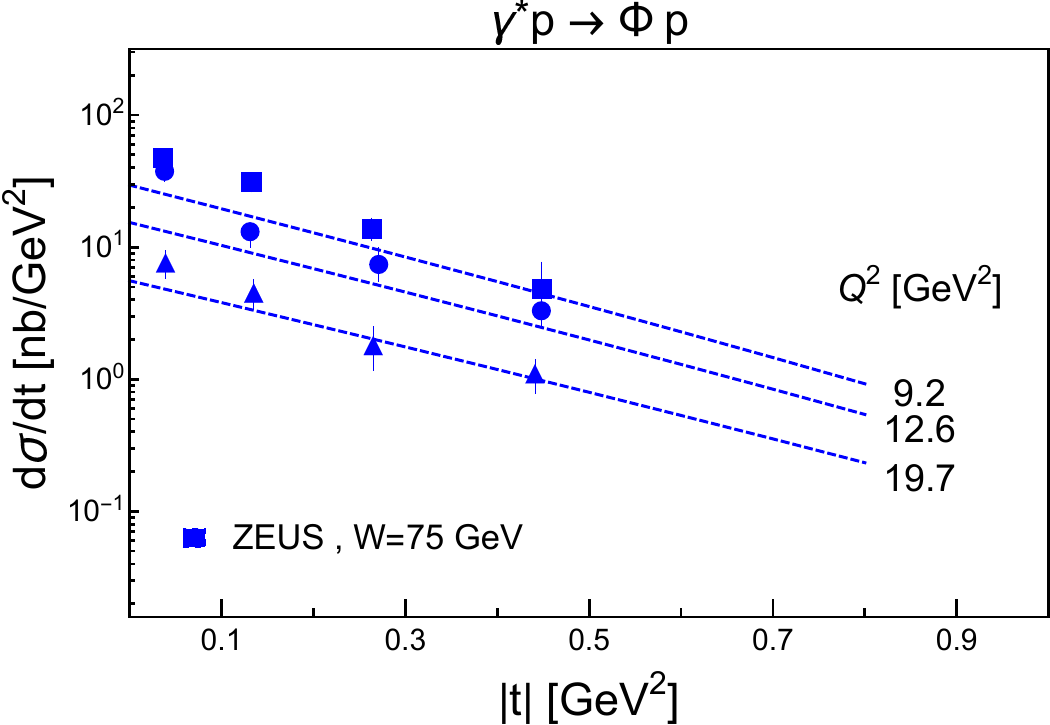}
  \end{subfigure}

  \vspace{1em}

  \begin{subfigure}{0.45\textwidth}
    \includegraphics[width=\linewidth]{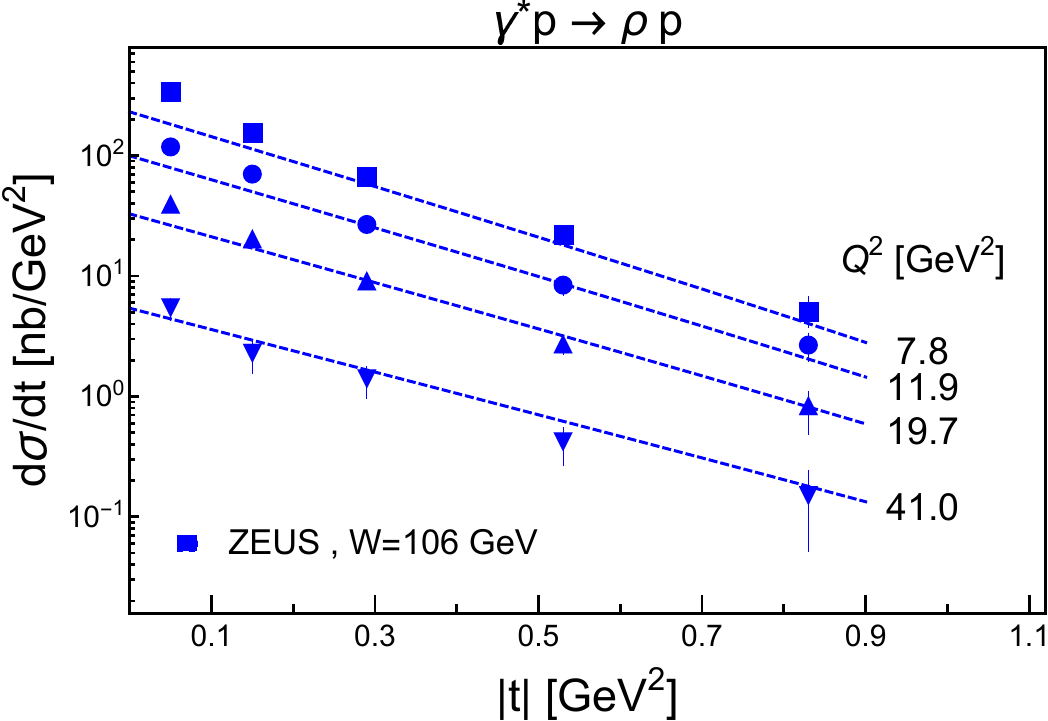}
  \end{subfigure}
  \hspace{0.5em}
  \begin{subfigure}{0.45\textwidth}
    \includegraphics[width=\linewidth]{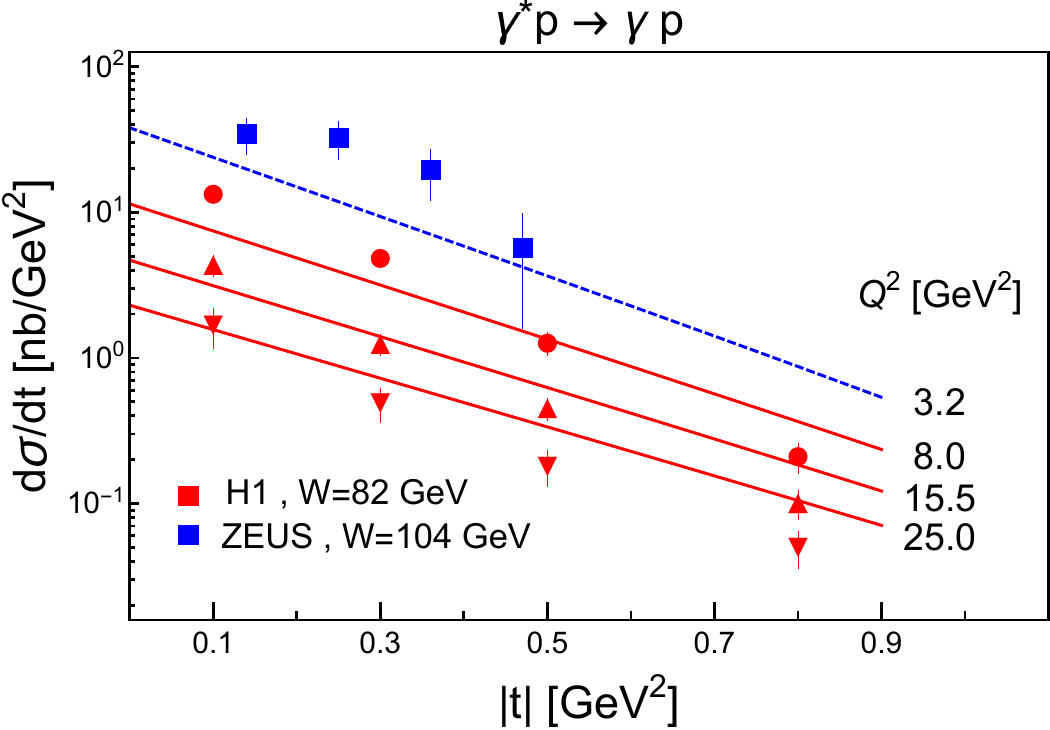}
  \end{subfigure}

  \caption{Differential cross sections for $J/\psi$, $\phi$, $\rho$ and $\gamma$, as a function of $t$. The red solid  and blue dashed curves are calculations to be compared with H1 and ZEUS data \cite{H1:2005dtp,ZEUS:2004yeh,ZEUS:2005bhf,ZEUS:2007iet,ZEUS:2008hcd,H1:2009wnw}.}
  \label{fig:sigt}
\end{figure}

 In Fig.~\ref{fig:sigt} we show the cross section as a function of $t$. One can also notice the agreement is better for larger $Q^2$ than small $Q^2$. The t-dependence of the curves are generally close to experiment data, except for the photon. This may be due to the over-simplified real photon $q\bar{q}$-LFWFs obtained with contact interaction model under DSEs approach. Employing realistic interaction model such as Maris-Tandy-like models can produce more realisitic photon $q\bar{q}$-LFWFs, and may bring the calculation closer to data. 

\section{Summary\label{sec:con}}
Supplementing with real photon and $\phi$ meson's DS-BSEs based $q\bar{q}$-LFWFs, we study the diffractive electroproduction of various light and heavy vector particles, i.e., the $\gamma$, $\rho$, $\phi$ and $J/\psi$. It is found that the light mesons host considerable higher Fock-states, as summarized in Table.~\ref{tab:N}. In particular, the $\phi$ meson, like the $\rho$ meson, remains far from where the $\langle q\bar{q}|q\bar{q}\rangle = 1$ approximation can be seriously taken. On the other hand, the $\phi$ meson $q\bar{q}$-LFWFs are quite different from $\rho$ meson's in profile, exhibiting $\phi$'s distinctive and unique properties among the vector mesons.

Bearing in mind that light mesons contain significant higher Fock-state components, we study exclusive vector meson production under the leading Fock-state truncation. Our key finding is shown in Fig.~\ref{fig:sigQ2}, i.e., while the $q\bar{q}$ truncated calculation agrees well with HERA  $J/\psi$ data for $Q^2$ as low as 0, the agreement for $\rho$, $\phi$ and $\gamma$ with HERA data only starts from a certain $Q_c^2\gtrsim$ 10 GeV$^2$. This is in line with the twist expansion idea within QCD factorization of exclusive processes. We emphasize that we do not introduce or tune any model parameter in this paper, hence the calculation demonstrates a robust prediction as a joint effort of color dipole approach and nonperturbative DS-BSEs study.

 This study also highlights the importance of diffractive $\phi$  electroproduction study. In color dipole model study, the lighter mesons have larger dipole size and are more sensitive to saturation effects. Yet as we show, they also push the $Q_c$ toward larger value where saturation  effect gets weakened. The $\phi$ meson thus balances the advantages between light and heavy vector mesons. We note that existing simulation of exclusive $\phi$ production in electron-nuclei collisions at electron-ion collider shows saturation can have visible effect in the domain of $Q^2\approx 10$ GeV$^2$ \cite{Toll:2012mb}. This result could be revisited with DS-BSEs based $q\bar{q}$-LFWFs. If it stays true, then a kinematic window is open for studying saturation effects in nuclei, based on a novel determination of leading Fock-state contribution to diffractive light vector particle electroproductions in presence of complex higher Fock-states.

\begin{acknowledgments}
Chao Shi thanks Yaping Xie and Xurong Chen for helpful discussions.
\end{acknowledgments}

\appendix

\bibliography{DVMP}

\end{document}